\documentclass{aastex61}
\usepackage{natbib}
\usepackage{amsmath}

\newcommand\aastex{AAS\TeX}

\received{}
\revised{}
\accepted{}
\submitjournal{ApJ}

\shorttitle{\aastex\ sample article}
\shortauthors{Yong Zhou, Zhi-Chao Zhao, Zhe Chang}

\begin{document}

\title{Searching for a cosmological preferred direction with 147 rotationally supported galaxies}

\correspondingauthor{Yong Zhou}
\email{zhouyong@ihep.ac.cn}

\author[0000-0002-3901-0228]{Yong Zhou}

\author{Zhi-Chao Zhao}

\author{Zhe Chang}

\affiliation{Institute of High Energy Physics, Chinese Academy of Sciences, Beijing 100049, China}
\affiliation{School of Physics Sciences, University of Chinese Academy of Sciences, Beijing 100049, China}

\begin{abstract}

It is well known that the Milgrom's MOND (modified Newtonian dynamics) explains well the mass discrepancy problem in galaxy rotation curves. The MOND predicts a universal acceleration scale below which the Newtonian dynamics is invalid yet. The universal acceleration scale we got from the SPARC dataset is $g_{\dag}=1.02\times10^{-10} \rm ~m~s^{-2}$. Milgrom suggested that the acceleration scale may be a fingerprint of cosmology on local dynamics and related with the Hubble constant $g_{\dag}\sim cH_0$. In this paper, we use the hemisphere comparison method with the SPARC dataset to investigate the spatial anisotropy on the acceleration scale. We find that the hemisphere of the maximum acceleration scale is in the direction $(l,b) = ({175.5^\circ}^{+6^\circ}_{-10^\circ}, {-6.5^\circ}^{+8^\circ}_{-3^\circ})$ with $g_{\dag,max}=1.10\times10^{-10} \rm ~m~s^{-2}$, while the hemisphere of the minimum acceleration scale is in the opposite direction $(l,b) = ({355.5^\circ}^{+6^\circ}_{-10^\circ}, {6.5^\circ}^{+3^\circ}_{-8^\circ})$ with $g_{\dag,min}=0.76\times10^{-10} \rm ~m~s^{-2}$. The maximum anisotropy level reaches up to $0.37\pm0.04$. Robust tests present that such a level of anisotropy can't be reproduced by a statistically isotropic data. In addition, we show that the spatial anisotropy on the acceleration scale has little correlation with the non-uniform distribution of the SPARC data points in sky. We also find that the maximum anisotropy direction is close with other cosmological preferred directions, especially the direction of the ``Australia dipole'' for the fine structure constant.

\end{abstract}




\section{Introduction} \label{sec:intro}

The cosmological principle states that the Universe is homogeneous and isotropic on large scale \citep{weinberg2008cosmology}. However, this principle seems to be faced with great challenges \citep{Perivolaropoulos:2008ud,Perivolaropoulos:2011hp}. On large scale, the cosmological anisotropy has been tested by physical observations for many times. For example, by investigating the peculiar velocities of the clusters of galaxies, physicists found that there exists a large-scale bulk flow \citep{Kashlinsky:2008ut,Kashlinsky:2009dw,Watkins:2008hf,Feldman:2009es}. This bulk flow points towards the direction $(l,b)=(282^\circ\pm11^\circ , 6^\circ\pm6^\circ)$. In this direction, the peculiar velocity reaches up to $400 \rm ~km~s^{-1}$ on scales of $100 ~h^{-1}\rm ~Mpc$, which is much larger than the value $110 \rm ~km~s^{-1}$ constrained by WMAP5 \citep{Dunkley:2008ie}. By analyzing the quasar absorption spectra, physicists found the variation of the fine structure constant $\alpha=e^2/c\hbar$ could be well represented by an angular dipole model pointing to the direction $(l,b)=(330^\circ\pm15^\circ, -13^\circ\pm10^\circ)$ with significance at the 4.2$\sigma$ confidence level \citep{Webb:2010hc,King:2012id}. In addition, the analysis about Union2 type Ia supernova data hints that the universe also has a privileged direction pointing to $(l, b) = ({309^\circ}^{+23^\circ}_{-3^\circ}, {18^\circ}^{+11^\circ}_{-10^\circ})$. In this direction, the universe expansion has the maximum acceleration \citep{Antoniou:2010gw,Cai:2011xs}. Furthermore, the anisotropy of cosmic microwave background (CMB) from the Planck satellite has been confirmed with significance at the 3$\sigma$ confidence level \citep{Ade:2013nlj}. And the low multipoles in the CMB angular power spectrum approximately point to a common direction \citep{Tegmark:2003ve,Bielewicz:2004en,Land:2005ad,Mariano:2012ia}. All these facts hint that there may exist a privileged direction in our universe.

Recently, McGaugh et al. \citep{Lelli:2016zqa,McGaugh:2016leg} employed the new extended Spitzer Photometry and Accurate Rotation Curves (SPARC) dataset, to investigate the radial acceleration relation in rotationally supported galaxies. They analyzed 2693 points in 147 galaxies and found out a fitting function with a unique parameter. This function shows a tight correlation between the centripetal acceleration traced by rotation curves and the baryonic (gravitational) acceleration predicted by the observed distribution of baryons. The correlation can hardly be explained by the dark matter hypothesis, but it can be predicted by the MOND theory \citep{Milgrom:2007br,McGaugh:2008nc,Milgrom:2016uye}, where the paremeter is a universal acceleration scale in the MOND.

The acceleration scale is a vitally important quantity in the MOND,  below which the Newtonian dynamics is invalid yet. Usually, the MOND predicts a universal acceleration scale for all galaxies \citep{Milgrom:1983ca,Milgrom:1983pn,Milgrom:2002tu,Milgrom:2008rv,Milgrom:2012xw}. However, in practice, physicists take the acceleration scale as a free parameter to fit the galaxy rotation curve, and they found that different galaxy may  have different acceleration scale \citep{Begeman:1991iy,Bottema:2002zv,Gentile:2010xt,Swaters:2010qe}. It inspires us to investigate the spatial anisotropy on the acceleration scale. In addition, the acceleration scale is independent with other physical parameter, so that the spatial anisotropy on the acceleration scale is probably related with the cosmological preferred direction. In fact, Milgrom \citep{Milgrom:1998ak} has suggested that cosmology is not simply an application of a relativistic version of MOND but a unit with it. The acceleration scale may be a fingerprint of cosmology on local dynamics and related with Hubble constant as $g_{\dag}\sim cH_0$.

In this paper, we make use of the hemisphere comparison method \citep{Schwarz:2007wf} with the SPARC dataset to investigate the spatial anisotropy on the acceleration scale. The anisotropy level is described by the normalized difference of the acceleration scale on two opposite hemispheres. The best fitting value of the acceleration scale on each hemisphere is obtained by using the orthogonal-distance-regression algorithm \citep{boggs1987stable,boggs1990orthogonal} on the McGaugh's function \citep{McGaugh:2016leg}. Then, we would find out the maximum anisotropy level and the corresponding direction from all selected directions. In order to check the validity of the maximum anisotropy level, we rebuild the SPARC dataset by statistically isotropic data, and examine whether the maximum anisotropy level from the SPARC dataset could be reproduced by this isotropic data. We also check the correlation between the spatial anisotropy on the acceleration
scale with the non-uniform distribution of the SPARC data points in sky.

The rest of this paper is organized as follows: In Sect.\ref{sec:SPARC}, we give a brief introduction to the SPARC dataset and the radial acceleration relation. In Sect.\ref{sec:Anisotropy}, we use the hemisphere comparison method to search for the maximum anisotropy level and corresponding direction on the SPARC dataset. Then we compare the anisotropy level with statistically isotropic data and the non-uniform distribution of the SPARC data points in sky. Finally, conclusions and discussions are given in Sect.\ref{sec:conclusion}.

\begin{figure*}
\begin{center}
\includegraphics[width=\textwidth]{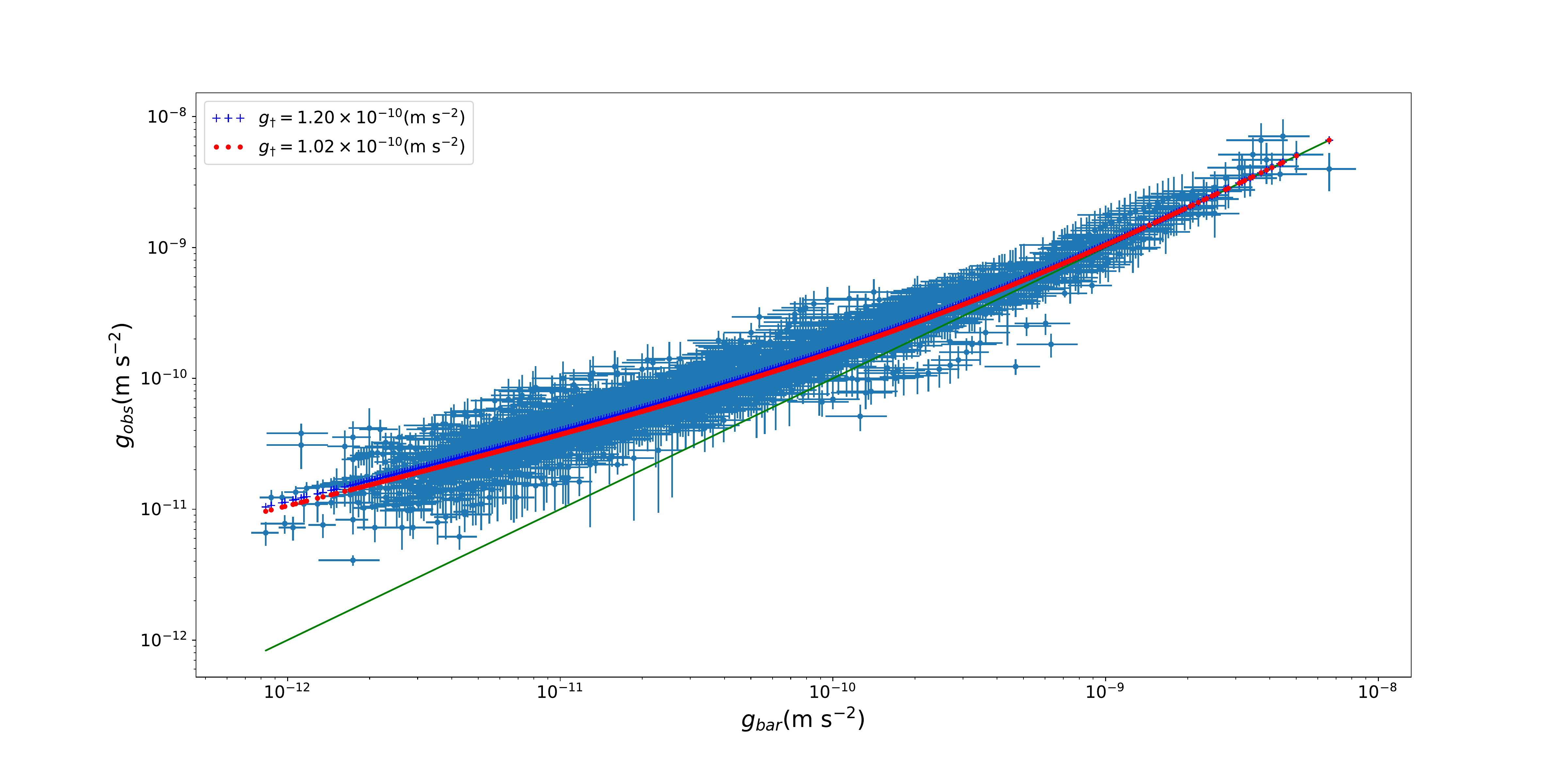}
\caption{The radial acceleration relation between the centripetal acceleration $g_{obs}$ and the baryonic acceleration $g_{bar}$ for all 2693 data points in 147 galaxies. The original acceleration is plotted in points with errorbars. The blue dotted plus line corresponds to the McGaugh's fitting curve with $g_{\dag}=1.20\times10^{-10}\rm ~m~s^{-2}$. The red dotted line corresponds to our fitting curve with $g_{\dag}=1.02\times10^{-10}\rm ~m~s^{-2}$. Two fitting curves are very close to each other. The solid line is the line of unity ($g_{obs}=g_{bar}$).}\label{fig:SPARC}
\end{center}
\end{figure*}

\section{The SPARC dataset and the radial acceleration relation} \label{sec:SPARC}

The SPARC dataset is a sample of 175 disk galaxies with new surface photometry at $3.6 \rm~\mu m$ and high-quality rotation curves from previous $\rm HI/H\alpha$ studies \citep{Lelli:2016zqa}. All these galaxies are rotationally supported galaxies. The SPARC spans a broad range of morphologies, luminosities, sizes and surface brightnesses. This is a good platform for investigating the radial acceleration relation. The SPARC dataset could be available on this website \footnote{\url{http://astroweb.cwru.edu/SPARC/}}.

For seeking the radial acceleration relation, a few modest quality criteria has been implemented to exclude some unreliable data. This operation finally leaves a sample of 2693 data points in 147 galaxies. More details could be found in reference \citep{McGaugh:2016leg}. Using these data, McGaugh et al. obtain a fitting function describing well these data. The fitting process employs an orthogonal-distance-regression algorithm that considers errors on both variables. The fitting function is of the form
\begin{eqnarray}\label{eq:nuy}
g_{obs}=\frac{g_{bar}}{1-e^{-\sqrt{g_{bar}/g_{\dag}}}},
\end{eqnarray}
where $g_{obs}$ is the observed centripetal acceleration traced by rotation curves and $g_{bar}$ is the baryonic (gravitational) acceleration predicted by the distribution of baryonic mass. There is a unique fitting parameter, which is the acceleration scale $g_{\dag}$. They found $g_{\dag}=\rm [1.20\pm0.02~(random)\pm0.24~(systematic)]\times10^{-10}~m~s^{-2}$. The original 2693 data points and the fitting curve are plotted in Fig.\ref{fig:SPARC}.

The fitting function \eqref{eq:nuy} has two limiting cases. In the Newton limit $g_{bar}\gg g_{\dag}$, the function \eqref{eq:nuy} becomes $g_{obs} \approx g_{bar}$ and the Newtonian dynamics is restored. In the deep-MOND limit $g_{bar}\ll g_{\dag}$, the function \eqref{eq:nuy} becomes $g_{obs} \approx \sqrt{g_{bar} g_{\dag}}$, where the mass discrepancy appears. This is consistent with the MOND theory \citep{Milgrom:1983ca,Milgrom:2016uye}.

The radial acceleration relation between the centripetal acceleration and the baryonic acceleration is tight for all 2693 data points in 147 galaxies. From Fig.\ref{fig:SPARC}, one can see clearly that a large majority of the data points are close to the fitting curve. However, for individual galaxies, the data points could have some deviation from the fitting curve. It implies that there may be other acceleration scales corresponding to these galaxies. Further, the spatial distribution of the acceleration scale may reflect the cosmological anisotropy.

The orthogonal-distance-regression algorithm \citep{boggs1987stable,boggs1990orthogonal} employed in fitting process is also used in this paper.  We define the chi-square as
\begin{eqnarray}\label{eq:chi}
\chi^2=\sum^n_{i=1}\frac{[g_{th}(g_{bar,i}+\delta_i,g_{\dag})-g_{obs,i}]^2}{\sigma^2_{obs,i}}+\frac{\delta_i^2}{\sigma^2_{bar,i}},
\end{eqnarray}
where subscript $i$ represent the $i$th data, and $n=2693$ is the total number of data points. $\sigma_{bar}$ and $\sigma_{obs}$ are the uncertainty of $g_{bar}$ and $g_{obs}$, respectively. $\delta_i$ is an auxiliary parameter for finding out the weighted orthogonal (shortest) distance for the $i$th data point from the curve $g_{th}(g_{bar},g_{\dag})$. The expression for the curve is same as the right-hand side of the function \eqref{eq:nuy}, i.e.
\begin{eqnarray}\label{eq:theoretical}
g_{th}(g_{bar},g_{\dag})=\frac{g_{bar}}{1-e^{-\sqrt{g_{bar}/g_{\dag}}}},
\end{eqnarray}
where $g_{th}$ represent the theoretical centripetal acceleration. Then the chi-square is the sum of the squares of the weighted orthogonal distances from the curve $g_{th}(g_{bar},g_{\dag})$ to the $n$ data points. Finally, we minimize the chi-square to find out the best fitting parameter for $g_{\dag}$.

Before we search for the maximum anisotropy direction from the SPARC dataset, we repeat the fitting process with the chi-square \eqref{eq:chi} for all 2693 data points in 147 galaxies. The best fitting value for the acceleration scale is $g_{\dag}=(1.02\pm0.02)\times10^{-10}\rm ~m~s^{-2}$. Here, we haven't account for the systematic uncertainty. This fitting curve is also plotted as the red dotted line in Fig.\ref{fig:SPARC}.

\section{Maximum anisotropy direction of the SPARC dataset} \label{sec:Anisotropy}

The hemisphere comparison method is widely adopted in searching for the cosmological anisotropy \citep{Antoniou:2010gw,Cai:2011xs,Lin:2016jqp}. We also adopt this method in the present study. One reason we choose this method is that it optimizes the statistics because there is a large number of data points in each hemisphere. The SPARC dataset doesn't include the galactic coordinate for each galaxy, and we complement them from the NED database \footnote{\url{http://ned.ipac.caltech.edu/}}. Together with the radial accelerations and its uncertainties in the SPARC dataset, the data is completed now for our study.

The hemisphere comparison method mainly consists of the following steps.
\begin{enumerate}
\item \label{itm:1}Generate an arbitrary direction with a unit vector $\hat{\textsl{\textbf{n}}}=\cos(b)\cos(l)\hat{\textsl{\textbf{i}}}+\cos(b)\sin(l)\hat{\textsl{\textbf{j}}}+\sin(b)\hat{\textsl{\textbf{k}}}~$ in the sky, where $l$ and $b$ are longitude and latitude, respectively, in the galactic coordinate system.
\item \label{itm:2}According to the sign of the inner product $\cos\theta_i=\hat{\textsl{\textbf{n}}}\cdot\hat{\textsl{\textbf{p}}}_i$, where $\hat{\textsl{\textbf{p}}}_i=\cos(b_i)\cos(l_i)\hat{\textsl{\textbf{i}}}+\cos(b_i)\sin(l_i)\hat{\textsl{\textbf{j}}}+\sin(b_i)\hat{\textsl{\textbf{k}}}$ is the unit vector pointing to the $i$th galaxy, spilt the dataset into two subsets. Thus, the hemisphere aligns with the direction of the unit vector $\hat{\textsl{\textbf{n}}}$ (defined as `up') corresponds to one subset, while the opposite hemisphere (defined as `down') corresponds to another subset.
\item \label{itm:3}Find the best fitting value of the acceleration scale $g_{\dag}$ on each hemisphere. Define the anisotropy level by the normalized difference
\begin{eqnarray}\label{eq:anisotropy}
D_{g_{\dag}}(l,b)=\frac{\Delta g_{\dag}}{\overline{g}_{\dag}}=2\frac{g_{\dag,u}-g_{\dag,d}}{g_{\dag,u}+g_{\dag,d}},
\end{eqnarray}
where $g_{\dag,u}$ and $g_{\dag,d}$ are the best fitting value in the `up' and `down' hemisphere, respectively.
\item \label{itm:4}Repeat steps \ref{itm:1} to \ref{itm:3} for sufficient directions and find the maximum value of $|D_{g_{\dag}}|$ as well as the corresponding direction $(l,b)$.
\end{enumerate}

\begin{figure*}
\begin{center}
\includegraphics[width=\textwidth]{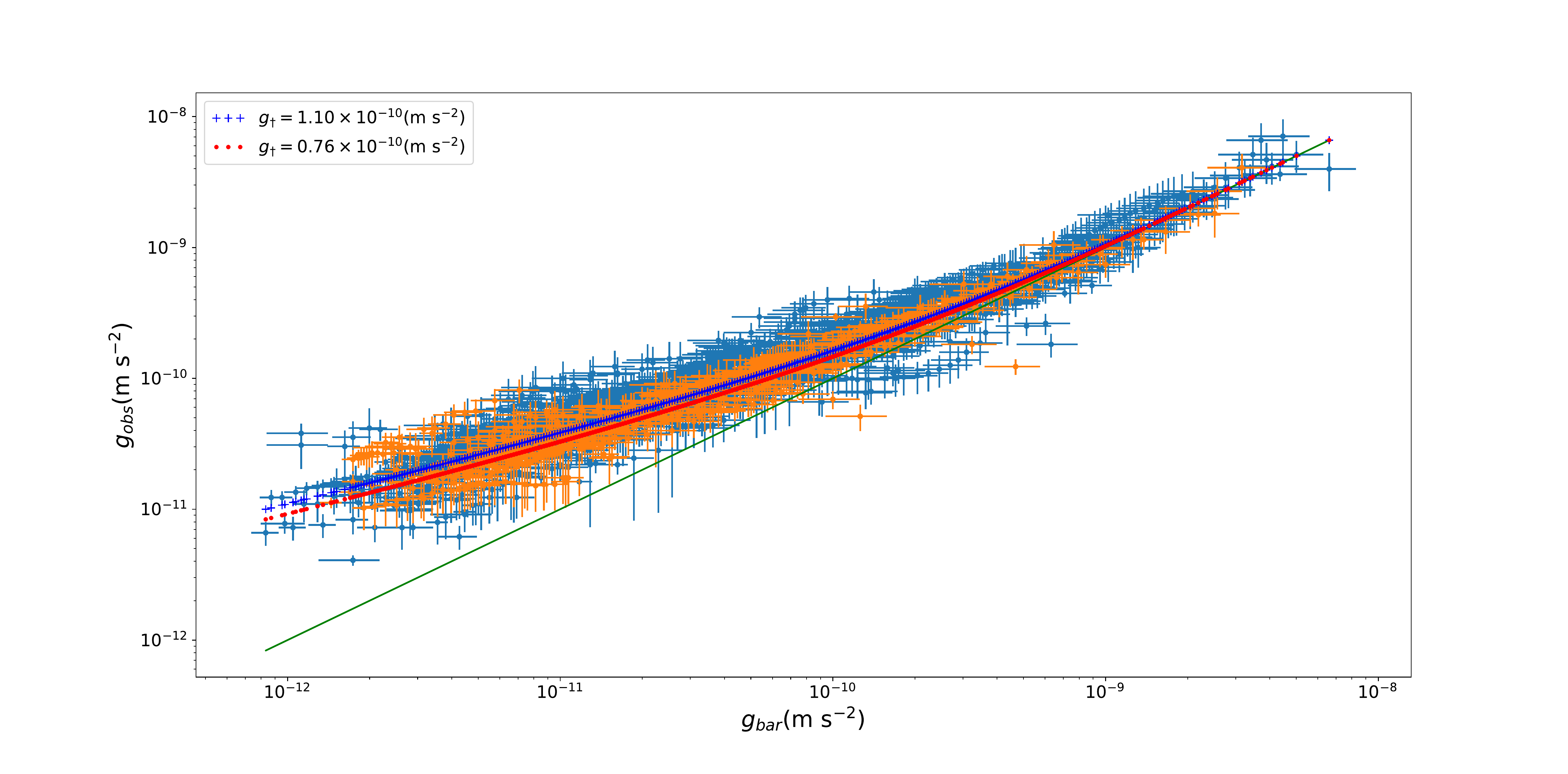}
\caption{The best fitting curves for two opposite hemispheres in the direction $(l,b)=(175.5^\circ,-6.5^\circ)$, which has a maximum anisotropy level. The blue dotted plus line corresponds to the maximum acceleration scale $g_{\dag,max}=1.10\times10^{-10}\rm ~m~s^{-2}$, which is the best fitting curve for the cyan data points in `up' hemisphere. The red dotted line corresponds to the minimum acceleration scale $g_{\dag,min}=0.76\times10^{-10}\rm ~m~s^{-2}$, which is the best fitting curve for the orange data points in `down' hemisphere.}\label{fig:SPARC_2}
\end{center}
\end{figure*}

In the step \ref{itm:3}, the best fitting value of the acceleration scale on each hemisphere is also obtained by the orthogonal-distance-regression algorithm. In order to find out the maximum anisotropy level, the number of directions must be large enough to ensure that the angular region corresponding to the maximum anisotropy level can be covered. Here, we divide the sky into $1^\circ\times 1^\circ$ grids, and choose the center of each grid as the direction of hemisphere. The total number of directions is 64800. Searching for all these directions, we find that the hemisphere of the maximum acceleration scale is towards the direction $(l,b)=(175.5^\circ,-6.5^\circ)$ with $g_{\dag,max}=(1.10\pm0.02)\times10^{-10}\rm ~m~s^{-2}$. While the hemisphere of the minimum acceleration scale is in the opposite direction $(l,b)=(355.5^\circ,6.5^\circ)$ with $g_{\dag,min}=(0.76\pm0.02)\times10^{-10}\rm ~m~s^{-2}$. The error in the acceleration scale is a $1\sigma$ value derived from the fitting process. The result is plotted in Fig.\ref{fig:SPARC_2}. Basing on the definition of anisotropy level in Eq.\eqref{eq:anisotropy}, we get the maximum anisotropy level, $D_{g_{\dag,max}}=0.37$. And the $1\sigma$ error for the maximum anisotropy level is propagated from the uncertainties of acceleration scale,
\begin{eqnarray}\label{eq:error}
\sigma_{D_{g_{\dag}}}=4\frac{\sqrt{g^2_{\dag,u}\sigma^2_{g_{\dag,d}}+g^2_{\dag,d}\sigma^2_{g_{\dag,u}}}}{(g_{\dag,u}+g_{\dag,d})^2}=0.04,
\end{eqnarray}
where the $\sigma_{g_{\dag}}$ is the $1\sigma$ value for the hemisphere acceleration scale. Thus, the maximum anisotropy level is $D_{g_{\dag},max}=0.37\pm 0.04$.

\begin{figure*}
\begin{center}
\includegraphics[width=\textwidth]{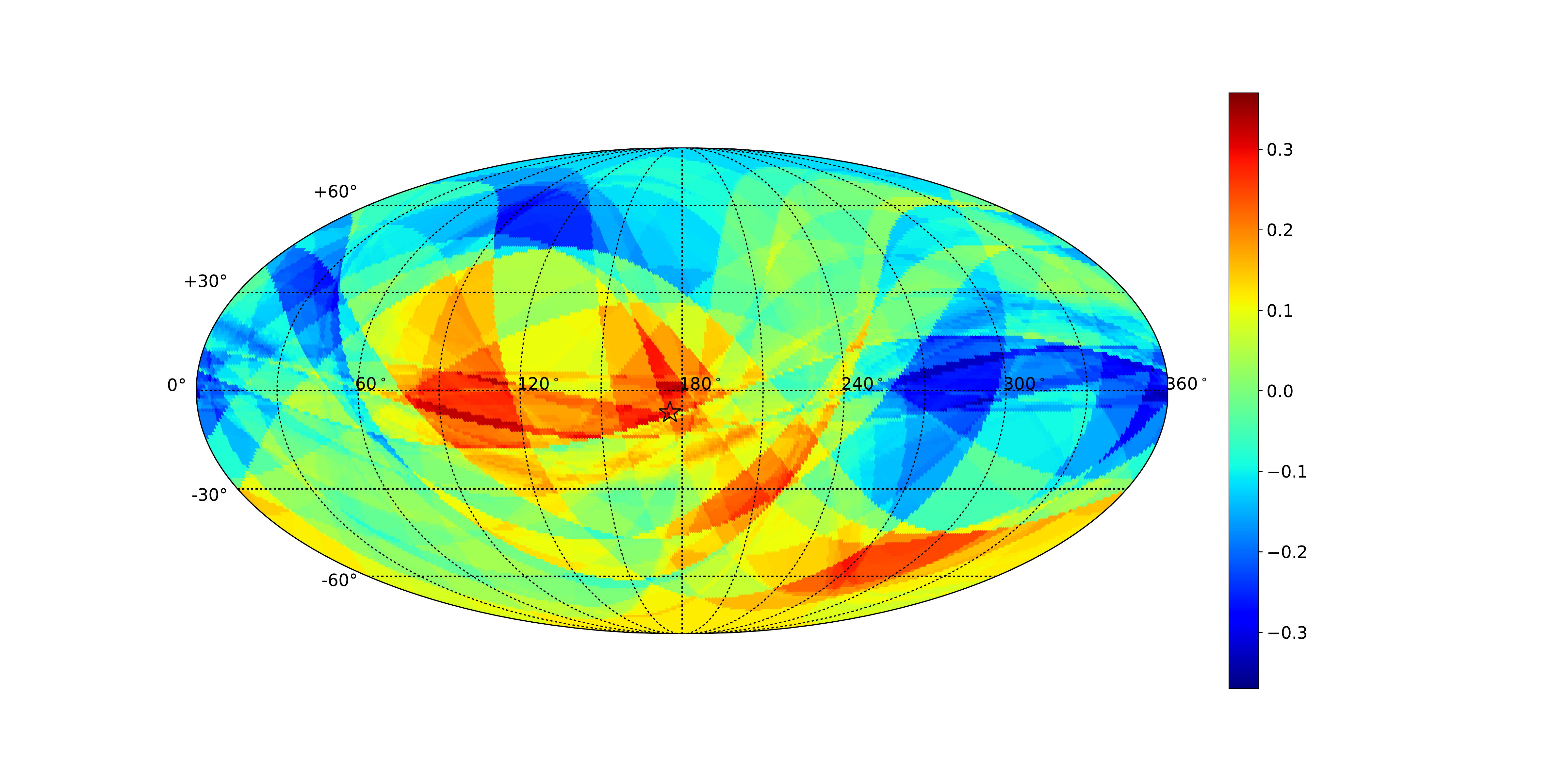}
\caption{The pseudo-colour map of the anisotropy level $D_{g_{\dag}}$ derived from the hemisphere comparison method. The maximum anisotropy level is $D_{g_{\dag},max}=0.37\pm 0.04$, which corresponds with the direction (pentagram) $(l,b) = ({175.5^\circ}^{+6^\circ}_{-10^\circ}, {-6.5^\circ}^{+8^\circ}_{-3^\circ})~(2\sigma)$. The submaximum anisotropy level is $D_{g_{\dag},sub}=0.34\pm 0.05$, which corresponds with the direction $(l,b) = ({114.5^\circ}^{+6^\circ}_{-33^\circ}, {2.5^\circ}^{+3^\circ}_{-14^\circ})~(1\sigma)$. Except for these two directions, anisotropy level is less than 0.3.}\label{fig:HCmethod}
\end{center}
\end{figure*}

To find out the confidence angular region of the maximum anisotropy direction, we pick up the direction that corresponds to an anisotropy level within $2\sigma$ from the maximum anisotropy level. Here, we assume that the maximum anisotropy level is Gaussian, and we choose $2\sigma$ instead of $1\sigma$ (there is only two directions located in $1\sigma$ region).  The final result about the maximum anisotropy direction (with $2\sigma$ random error) is
\begin{eqnarray}\label{eq:up}
(l,b) = ({175.5^\circ}^{+6^\circ}_{-10^\circ}, {-6.5^\circ}^{+8^\circ}_{-3^\circ}),
\end{eqnarray}
or equivalently, the minimum anisotropy direction is its opposite direction
\begin{eqnarray}\label{eq:down}
(l,b) = ({355.5^\circ}^{+6^\circ}_{-10^\circ}, {6.5^\circ}^{+3^\circ}_{-8^\circ}).
\end{eqnarray}
This results are plotted in Fig.\ref{fig:HCmethod}. Except the maximum anisotropy direction, we find that there have another direction $(l,b) = ({114.5^\circ}^{+6^\circ}_{-33^\circ}, {2.5^\circ}^{+3^\circ}_{-14^\circ})~(1\sigma)$, in this direction the anisotropy level is up to $D_{g_{\dag},sub}=0.34\pm 0.05$. At other direction, the anisotropy level is less than 0.3.

\begin{figure}
\begin{center}
\includegraphics[width=8cm]{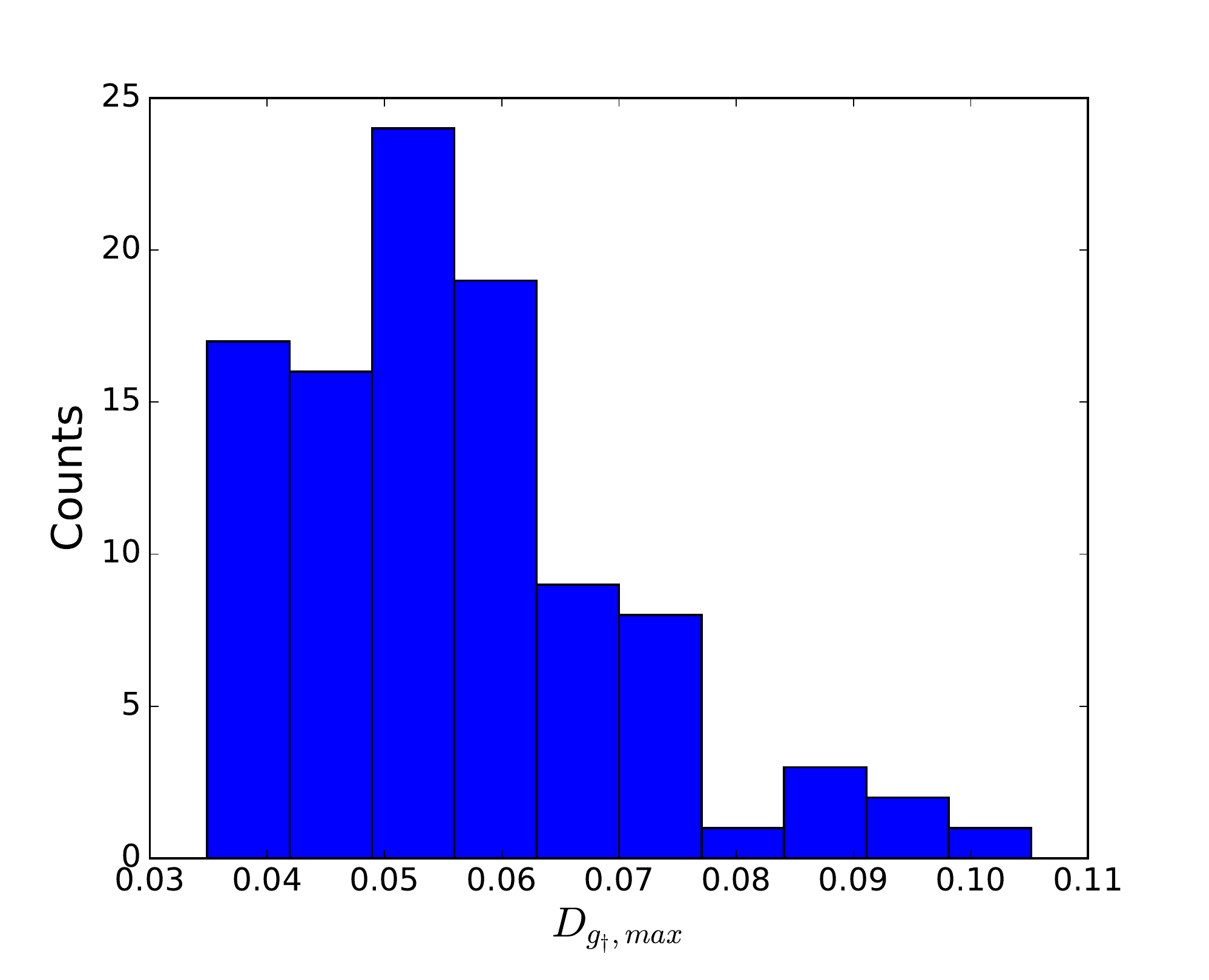}
\caption{The histogram of the maximum anisotropy level (i.e. $D_{g_{\dag},max}$) from the statistically isotropic data in the 100 simulations. The maximum anisotropy level is all around $D_{g_{\dag},max}=0.05\pm 0.02$.}\label{fig:Random}
\end{center}
\end{figure}

As a robust tests, we examine whether the maximum anisotropy level is consistent with the statistical isotropy. We rebuild the SPARC dataset by statistically isotropic data. The baryonic acceleration $g_{bar}$ and its uncertainty $\sigma_{bar}$ remains unchanged. The centripetal acceleration $g_{obs}$ has been replaced by a random number with a Gaussian distribution with the central value determined by the  theoretical centripetal acceleration $g_{th}$ in Eq.\eqref{eq:theoretical}, here $g_{\dag}=1.02\times 10^{-10}\rm ~m~s^{-2}$ is the universal acceleration scale, and the standard deviation is the same as the uncertainty $\sigma_{obs}$ of the centripetal acceleration $g_{obs}$. For generality, we then repeat the hemisphere comparison method for this statistical isotropic data 100 times. Because the centripetal acceleration is made up of random value, the hemisphere comparison result is different for each time. The result is plotted in Fig.\ref{fig:Random}. Even though the distribution of the anisotropy level seems to have some structures, but the maximum anisotropy level for each time is all around the value $D_{g_{\dag},max}=0.05\pm 0.02$. The maximum of the maximum anisotropy level is $D_{g_{\dag},max}=0.10\pm0.03$, which is fainter than that from the SPARC dataset. It means that the maximum anisotropy level from the original SPARC dataset couldn't be reproduced by such statistically isotropic data.

\begin{figure*}
\begin{center}
\includegraphics[width=\textwidth]{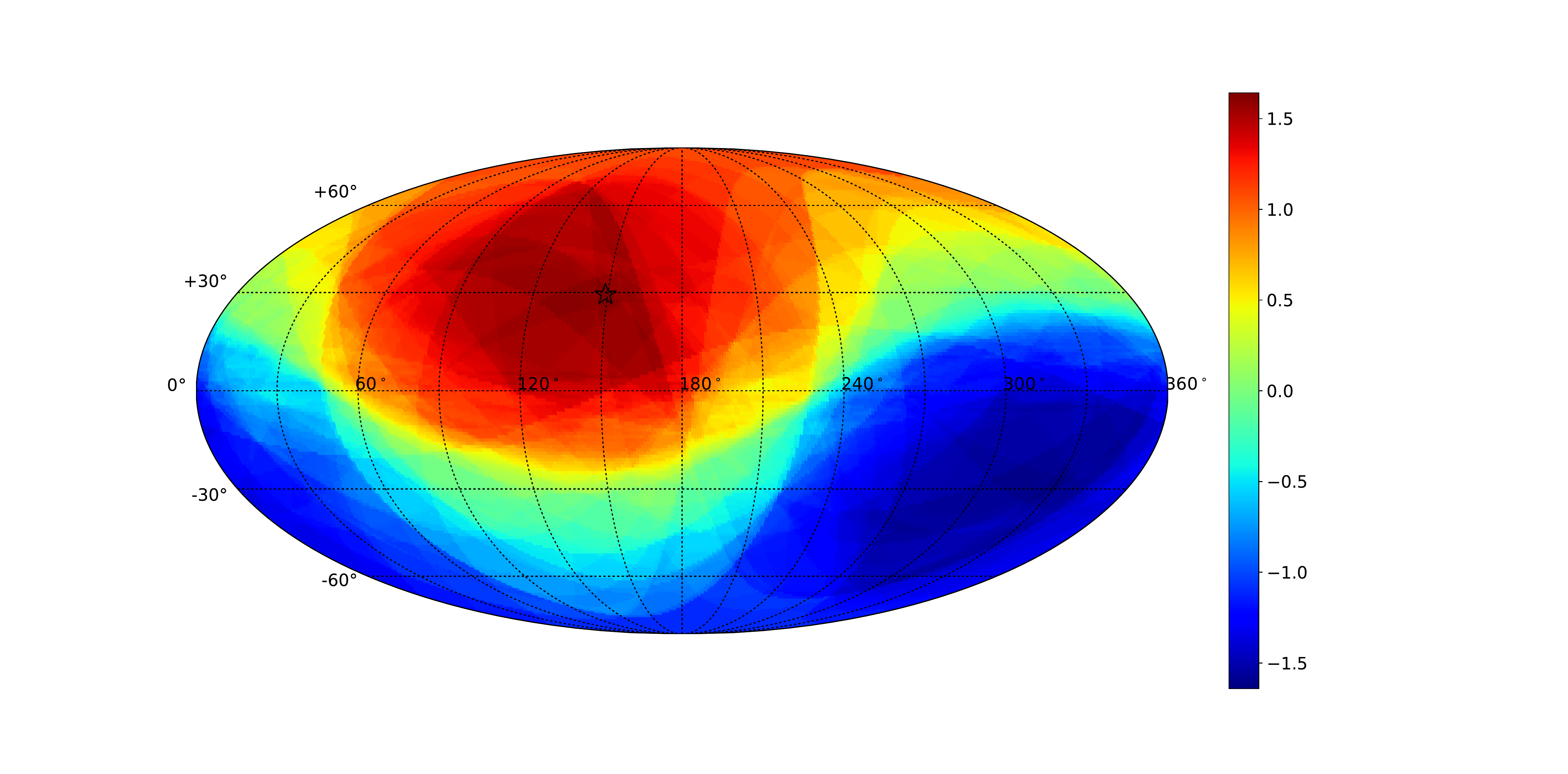}
\caption{The pseudo-colour map of the number difference of the SPARC data points between two opposite hemispheres. The most clustered direction (pentagram) is towards $(l,b)=(149.1^\circ,29.4^\circ)$, while the most sparse direction is in its opposite direction  $(l,b)=(329.1^\circ,-29.4^\circ)$.}\label{fig:Num}
\end{center}
\end{figure*}

Another worth discussing issue is that whether the maximum anisotropy level from the SPARC dataset is coming from the non-uniform distribution of the data points in sky. In the study of the anisotropy on supernova, it was pointed out that the anisotropy could be originated from the non-uniform distribution of the data points \citep{Chang:2014nca,Jimenez:2014jma}. Here, we take a similar investigation for the SPARC dataset. In order to trace the correlation between the anisotropy level on the acceleration scale and the distribution of the data points, we repeat the hemisphere comparison method again, but here we calculate the number difference of the SPARC data points in two opposite hemispheres. The direction we choose here is same as that in Fig.\ref{fig:HCmethod}. For a given direction $(l,b)$, define the number deffernce as
\begin{eqnarray}\label{eq:number}
D_N(l,b)=\frac{\Delta N}{\overline{N}}=2\frac{N_u-N_d}{N_u+N_d},
\end{eqnarray}
where $N_u$ and $N_d$ are the numbers of the SPARC data points in the `up' and `down' hemisphere, respectively. The distribution of $D_N(l,b)$ is shown in Fig.\ref{fig:Num}. As we can see clearly, the most clustered direction is towards $(l,b)=(149.1^\circ,29.4^\circ)$, while the most sparse direction is its opposite direction $(l,b)=(329.1^\circ,-29.4^\circ)$. In the most clustered direction, the `up' hemisphere have 2452 data points while the 'down' hemisphere have only 241 data points. The corresponding maximum number difference is $D_{N,max}=1.64$. In the direction of the maximum anisotropy level on the acceleration scale, we found the number of data points in `up' and `down' hemisphere are 2173 and 520, respectively. The corresponding number difference is $D_{N}=1.23$. Comparing the direction on the maximum anisotropy level and the maximum number difference, we found this two directions have a large angular separation $\Delta\theta\approx44^\circ$. In addition, the distribution of $D_{g_{\dag}}$ in Fig.\ref{fig:HCmethod} and $D_N$ in Fig.\ref{fig:Num} have large differences in overall outline. This implies the spatial anisotropy on the acceleration scale has little correlation with the non-uniform distribution of the SPARC data points in sky.

\section{Conclusions and Discussions} \label{sec:conclusion}

In this paper, we employed the hemisphere comparison method with the SPARC dataset to search the maximum anisotropy direction, which is probably related with the cosmological preferred direction. The anisotropy level is described by the normalized difference of the acceleration scale on two opposite hemispheres. Repeating for 64800 directions on the centre of each grid, we found that the maximum anisotropy level is $D_{g_{\dag},max}=0.37\pm 0.04$, which corresponds with the direction $(l,b) = ({175.5^\circ}^{+6^\circ}_{-10^\circ}, {-6.5^\circ}^{+8^\circ}_{-3^\circ})~(2\sigma)$.  We also found that the submaximum anisotropy level is $D_{g_{\dag},sub}=0.34\pm 0.05$, which corresponds with the direction $(l,b) = ({114.5^\circ}^{+6^\circ}_{-33^\circ}, {2.5^\circ}^{+3^\circ}_{-14^\circ})~(1\sigma)$. At other direction, the anisotropy level is not significant and less than 0.3. We then rebuilt the SPARC dataset by statistically isotropic data, and found that the maximum anisotropy level from the SPARC dataset couldn't be reproduced by such isotropic data. Finally, we pointed out that there is little correlation between the spatial anisotropy on the acceleration scale and the non-uniform distribution of the SPARC data points in sky.

As discussed in the introduction, the cosmological preferred directions have been reported by a series of independent cosmological observations. These directions are all located in an angular region less than a quarter of the North Galactic Hemisphere. In this paper, we found a possible preferred direction from the SPARC dataset. This direction is close to the above angular region. Especially, we found the preferred direction is close to the direction of a dipole model which describes the variation of the fine structure constant, and these two directions only have an angular separation of $\Delta\theta\approx32^\circ$. The consistency of these directions may hint that our universe is indeed anisotropic, and it could be caused by an underlying physical effect, such as the space-time anisotropy.

If the cosmological preferred direction is confirmed, the standard $\Lambda$CDM model should be modified. However, inevitably, the researches on the cosmological preferred direction still have a large of uncertainties. This research is in the same situation. The possible uncertainty partly comes from the original data. As we presented in Sect.\ref{sec:SPARC}, the baryonic acceleration is predicted by the distribution of baryonic mass, while the baryonic mass is converted by the the near-infrared luminosity. For different galaxy, their mass-to-light ratio could be different too. In the reference \citep{McGaugh:2016leg}, McGaugh et al. adopted a uniform mass-to-light ratio for all galaxies. They emphasized that it could only have slight impact on the fitting result for all data points, but it may impact the result of the hemisphere comparison method. The maximum anisotropy level may increase, or reduce. Another possible uncertainty comes from the deficiency of the galaxy, and most data points cluster in same direction. For the future researches on the anisotropy with the galaxy rotation curves, the dataset need to be extended and cover better the sky uniformly. By then the maximum anisotropy direction could be more convincing.

\acknowledgments
The authors are grateful to Dr. Hai-Nan Lin for useful discussions. They are also thankful for the open access of the SPARC dataset. This work is supported by the National Natural Science Fund of China under Grants No.11375203, No.11675182 and No.11690022.

\end{document}